\documentclass[twocolumn,aps,superscriptaddress,showpacs]{revtex4}

\usepackage{amssymb}
\usepackage{amsmath}
\usepackage{mathrsfs}
\usepackage{graphicx}
\usepackage{subfigure}
\usepackage[normalem]{ulem}
\usepackage[dvips]{color}

\begin{document}

\title{Properties of quark-matter cores in massive hybrid stars}
\author{He Liu}
\email{liuhe@qut.edu.cn}
\affiliation{Science School, Qingdao University of Technology, Qingdao 266000, China}
\affiliation{The Research Center of Theoretical Physics, Qingdao University of Technology, Qingdao 266033, China}
\author{Xiao-Min Zhang}
\email{zhangxm@mail.bnu.edu.cn}
\affiliation{Science School, Qingdao University of Technology, Qingdao 266000, China}
\affiliation{The Research Center of Theoretical Physics, Qingdao University of Technology, Qingdao 266033, China}
\author{Peng-Cheng Chu}
\email{kyois@126.com}
\affiliation{Science School, Qingdao University of Technology, Qingdao 266000, China}
\affiliation{The Research Center of Theoretical Physics, Qingdao University of Technology, Qingdao 266033, China}
\date{\today}
\begin{abstract}
 Using the constraints from astrophysical observations and heavy-ion experiments, we investigate the equation of state (EOS) of hybrid star matter and the properties of quark-matter cores in hybrid stars. The quark matter interactions in hybrid stars are described based on 3-flavor Nambu-Jona-Lasinio model with various vector and vector-isovector coupling constants. In this work, we find that the hybrid star matter EOS is more sensitive to the strength of the vector interaction, and the EOS becomes stiffer with increasing vector strength $R_V$. The vector-isovector interaction characterized by the coupling constant $R_{IV}$ make main contribution to the hadron-quark mixed phase. Meanwhile, we note that a step change of both the sound velocity and the polytropic index $\gamma$ occurs in the hadron-quark phase transition, and it is restored with the decrease of nucleon and lepton degrees of freedom in the high density quark phase. The approximate rule that the polytropic index $\gamma\leq 1.75$ can also be used as a criterion for separating hadronic matter from quark matter in our work.  Using the hybrid star matter EOS, we predict the radius and mass information of quark-matter cores inside hybrid stars. Although the coupling constants increase the hybrid star maximum mass up to $2.08M_{\odot}$, they also decrease the mass and radius of the quark core and the mixed core. With different quark coupling constants, we also find that the maximum mass and radius of the quark matter core in a stable hybrid star can reach $0.80M_{\odot}$ and 6.95 km, which are close to half of the maximum mass and radius of the complete star. However, properties of quark matter have no effect on the $M = 1.4M_{\odot}$ hybrid star as a result of no quark matter inner core, which can also be confirmed by the criterion of the polytropic index, and thus our results also indicate that the quark interactions have no effect on the tidal deformability $\Lambda_{1.4}$ of hybrid stars.
\end{abstract}


\maketitle
\section{Introduction}
\label{INT}
Studying  the equation of state (EOS) of strongly interacting matter is one of the main scientific goals of nuclear physics and astrophysics~\cite{Lat04,Ste05}. Researches based on astrophysical observations of neutron stars~\cite{Lat00,Cxu10,Wat16,Oer17,Bom18} have already led to significant constraints on the EOS of nuclear matter around but mostly below the saturation density ($\rho_0 \approx 0.16$ fm$^{-3}$). Although the EOS of strongly interacting matter at densities $2\rho_0 < \rho < 5\rho_0$ has also been constrained by the measurements of collective flows~\cite{Dan02} and subthreshold kaon production~\cite{Fuc06} in relativistic heavy-ion collisions, the lack of accurate first-principles predictions at supra-saturation density has so far prevented determination of the phase of matter inside neutron star~\cite{Ann20}. In recent years, there have been some inspiring progress in astrophysical observations on neutron stars, including the accurate mass determination of the massive object PSR J1614-2230 ($1.908\pm0.016M_{\odot}$)~\cite{Dem10,Fon16,Arz18}  and PSR J0348+0432 ($2.01\pm0.04M_{\odot}$)~\cite{Ant13}. The first simultaneous measurements of the mass and radius of a neutron star using the Neutron Star Interior Composition Explorer (NICER) data were those of the millisecond pulsar PSR J0030+0451. The two independent analyses predict ($68\%$ credible interval) $M=1.34^{+0.15} _{-0.16} M_{\odot}, R=12.71^{+1.14}_{-1.19}$ km~\cite{Ril19} and $M=1.44^{+0.15}_{-0.14} M_{\odot}, R=13.02^{+1.24} _{-1.06}$ km~\cite{Mil19}. More recently, the measurement based on NICER and X-ray Multi-Mirror (XMM-Newton) found that gravitational mass of PSR J0740+6620 is declared as $2.08\pm0.07M_{\odot}$, which is considered as the highest reliably determined neutron star mass~\cite{Cro20,Ril21,Mil21}. Its radius were determined with the results $12.39^{+1.30}_{-0.98}$ ~\cite{Ril21} and $13.71^{+2.61}_{-1.50}$ km~\cite{Mil21} ($68\%$  credible interval). And the radius range that spans the $\pm1\sigma$ credible intervals of all the radius estimates in the different frameworks is $12.45\pm 0.65$ km for a canonical mass $M=1.4M_{\odot}$ neutron star~\cite{Mil21}. The gravitational wave events GW170817~\cite{Abb17} and GW190814~\cite{Abb20} have provided more additional constraints on the EOS of the neutron star matter. The analysis of GW170817 by the LIGO/Virgo Collaboration has found with a $90\%$ confidence that the tidal deformability of the merging compact stars constrained as the range $70 <\Lambda_{1.4}<580$~\cite{Abb18}. Besides, the newly discovered compact binary merger GW190814 which has a secondary component of mass $(2.50 \sim 2.67) M_{\odot}$ at $90\%$ credible level has also aroused lots of debates on whether the candidate for the secondary component is a neutron star or a light black hole. There are thus two restrictive observational constraints on the stiffness of the EOS of the neutron star matter: EOSs that are too soft are eliminated by the discovery of massive neutron stars, whereas EOSs that are very stiff are inconsistent with the merger gravitational wave signal from the binary star merger GW170817, which disfavors large tidal deformation (and large radii)~\cite{Alf19}. The conclusions we listed before might be a signal of the existence of non-nucleonic degrees of freedom at high densities in neutron star matter, such as quark matter. Some recent studies, for instance, have also shown that quark-matter cores can appear in massive neutron stars~\cite{Ann20} and the presence of a first-order phase transition from hadronic matter to quark matter can imprint signatures in binary mergers observations~\cite{Alf19,Bau19}, as well as the binary-neutron-star (BNS) merger simulations indicate that the sudden decrease in the gravitational-wave frequency is closely related to the hadron-quark phase transition~\cite{Hua22}. Thus, the above observation events/objects comprise the multi-messenger data set for our following analysis on quark matter inside neutron stars.

QCD effective models that incorporate important properties and symmetries of the strong interaction are widely used for describing quark degrees of freedom in neutron star matter. It has been found that large mass constraints have been used to understand the properties of the hadron-quark phase transition as well as the EOS of quark matter in neutron stars with different QCD effective models (see, e.g., Refs.~\cite{Sho03,Alf13,Chu14,Bay16,Chu17,Han19,Alv19,Zha21}).  The Nambu-Jona-Lasinio (NJL) model is one of such effective models that considers chiral symmetry preserving interactions~\cite{Hat94,Bub05}, and that has been used to study the possible existence of quark matter inside neutron stars~\cite{Len12,Bal13,Mas13,Sha13,Xia13,Chu16}. Although the scalar part of the quark matter interaction in the NJL model can be constrained by the lattice QCD calculations~\cite{Kar01,Kar02}, its vector part or especially the isovector part, which is relevant to the properties of quark matter at high net baryon densities and isospin asymmetries, remains poorly understood. However, a recent study using the transport approach based on a 3-flavor NJL model has shown that the strength of the vector interaction can be extracted from the relative elliptic flow ($v_2$) difference between protons and antiprotons as well as between $K^+$ and $K^-$ in relativistic heavy-ion collisions measured in experiments from the beam-energy scan (BES) program~\cite{Jxu14}.  Again, the isovector couplings lead to different potentials of $u$ and $d$ quarks in isospin asymmetric quark matter, and thus server as a possible explanation of the elliptic flow splitting between $\pi^+$ and $\pi^-$ in relativistic heavy-ion collisions~\cite{Liu19}. It is thus of great interest to investigate properties of quark-matter cores inside neutron stars by employing the 3-flavor NJL model with vector and isovector couplings.

 In the present study, the neutron stars could be converted to hybrid stars with the hadron-quark phase transition. Using the constraints from astrophysical observations and heavy-ion experiments, the EOS of hybrid star matter and the properties of quark-matter cores in hybrid stars are investigated. We describe strange quark matter in hybrid stars based on the 3-flavor NJL model with vector and isovector couplings, as well as nuclear matter using an improved isospin- and momentum-dependent interaction (ImMDI) model. The ImMDI model is constructed from fitting cold nuclear matter properties at saturation density and the empirical nucleon optical potential~\cite{Jxu15,Jxu19}, and it has been extensively used in intermediate energy heavy-ion reactions to study the properties of asymmetric nuclear matter. The Gibbs construction~\cite{GLe92, GLe01} is adopted for the description of hadron-quark mixed phase, where the coexisting hadronic and quark phases need to satisfy the $\beta$-equilibrium and charge-neutral conditions. This article is organized as follows. In Sec. II, we describe the NJL model with  vector and isovector couplings for the quark matter. The properties of quark-matter cores inside hybrid stars, such as the EOS of hybrid star matter, the sound velocity, the polytropic index $\gamma$, the mass-radius relation and the dimensionless tidal deformability, is presented in Sec. III.  Section VI is devoted to the conclusions.

\section{The theoretical model }
\label{MODEL}
The hybrid EoS consists of a hadronic phase connected to a quark phase through a hadron-quark mixed phase. The possible appearance of hyperons is neglected, which is due to the fact that there are still large uncertainties on the hyperon-nucleon ($YN$) and hyperon-hyperon ($YY$) interactions in the nuclear medium~\cite{Hiy20,Con16}. Besides, following the results from Ref.~\cite{Jxu10}, the fraction of hyperons disappears quickly in hadron-quark mixed phase, which means that the effect of hyperons at super-saturation density, especially in the hadron-quark mixed phase, is expected to be small. Thus, we mainly focus on, in this work, the properties of quark-matter cores in hybrid stars without hyperons.

For quark matter, the lagrangian of the 3-flavor NJL model with vector and isovector interactions can be expressed as~\cite{Liu22}
\begin{eqnarray}
\mathcal{L}_{\textrm{NJL}} &=& \bar{q}(i\rlap{\slash}\partial-\hat{m})q
+\frac{G_S}{2}\sum_{a=0}^{8}[(\bar{q}\lambda_aq)^2+(\bar{q}i\gamma_5\lambda_aq)^2]
\notag\\
&-&K[det\bar{q}(1+\gamma_5)q+det\bar{q}(1-\gamma_5)q]
\notag\\
&-&\frac{G_V}{2}\sum_{a=0}^{8}[(\bar{q}\gamma_\mu\lambda_aq)^2+
(\bar{q}\gamma_5\gamma_\mu\lambda_aq)^2]
\notag\\
&+&G_{IS}\sum_{a=0}^{3}[(\bar{q}\lambda_aq)^2+(\bar{q}i\gamma_5\lambda_aq)^2]
\notag\\
&-&G_{IV}\sum_{a=0}^{3}[(\bar{q}\gamma_\mu\lambda_aq)^2+
(\bar{q}\gamma_5\gamma_\mu\lambda_aq)^2],
\end{eqnarray}
where $q = (u, d, s)^T$ and $\hat{m} = diag(m_u,m_d,m_s)$ are the quark fields and the current quark mass matrix with three flavors, respectively; $\lambda_a$ are the flavor SU(3) Gell-Mann matrices with $\lambda_0 = \sqrt{2/3}I$; $G_S$ and $G_V$  are the scalar and vector coupling constant, respectively; as well as the $K$ term represents the six-point Kobayashi-Maskawa-t'Hooft (KMT) interaction that breaks the axial $U(1)_A$ symmetry~\cite{Hoo76}. The additional $G_{IS}$ and $G_{IV}$ terms represent the scalar-isovector and the vector-isovector interactions, respectively. The Gell-Mann matrices with $a = 1, 2, 3$ in the two isovector interaction term are identical to the Pauli matrices in $u$ and $d$ space, and thus the isovector coupling terms break the SU(3) symmetry while keeping the isospin symmetry. In the present study, we employ the parameters $m_u = m_d = 3.6$ MeV, $m_s = 87$ MeV, $G_S\Lambda^2 = 3.6$, $K\Lambda^5 = 8.9$, and the cutoff value $\Lambda = 750$ MeV in the momentum integral given in Refs.~\cite{Bra13,Bub05}.

In the mean-field approximation, the energy density $\epsilon_Q$ of quark matter in detail can be written as
\begin{eqnarray}
\varepsilon_Q &=&-2N_c\sum_{i=u,d,s}\int_0^\Lambda\frac{d^3p}{(2\pi)^3}
E_i(1-f_i-\bar{f}_i)
\notag\\
&+&G_S(\sigma_u^2+\sigma_d^2+\sigma_s^2)+G_V(\rho_u^2+\rho_d^2+\rho_s^2)
\notag\\
&-&4K\sigma_u\sigma_d\sigma_s+G_{IS}(\sigma_u-\sigma_d)^2
\notag\\
&-&G_{IV}(\rho_u-\rho_d)^2-\varepsilon_{\textrm{vac}},
\end{eqnarray}
where the factor $N_c=3$ represents the color degeneracy of quark, $f_i$ and $\bar{f}_i$ are respectively the fermi distribution functions of quark and antiquark with flavor $i$. $\sigma_i$ and $\rho_i$ stand for the quark condensate and the net quark number density, respectively~\cite{Liu16,Liu20}; $E_i(p)=\sqrt{p^2+M_i^2}$ is the single quark energy; and $\varepsilon_{\textrm{vac}}$ is introduced to ensure $\varepsilon_Q=0$ in vacuum. The pressure at zero temperature can be given as
\begin{eqnarray}
P_Q=\sum_{i=u, d, s}\mu_i\rho_i-\varepsilon_Q.
\end{eqnarray}

In the quark phase, the system is composed of a mixture of quarks ($u$, $d$, and $s$) and leptons ($e$ and $\mu$) under the charge neutrality condition
\begin{equation}
\frac{2}{3}\rho_u-\frac{1}{3}(\rho_d+\rho_s)-\rho_e-\rho_\mu=0,
\end{equation}
and the $\beta$-equilibrium condition
\begin{eqnarray}
\mu_s&=&\mu_d=\mu_u+\mu_e,
\\
\mu_\mu&=&\mu_e.
\end{eqnarray}
In terms of the electron mass $m_e=0.511$MeV and the muon mass $m_\mu=106$ MeV, the lepton contributions to the energy density and the pressure are
\begin{eqnarray}
\varepsilon_L&=&\sum_{i=e, \mu}\frac{1}{\pi^2}\int_0^{p_f^i}\sqrt{p^2+m_i^2}p^2dp,
\\
P_L&=&\sum_{i=e, \mu}\mu_i\rho_i-\varepsilon_L,
\end{eqnarray}
where $p_f^i=(3\pi^2\rho_i)^{\frac{1}{3}}$ is the lepton fermi momentum. The total energy density and pressure including the contributions from both quarks and leptons in quark phase
are given by
\begin{eqnarray}
\varepsilon^Q&=&\varepsilon_Q+\varepsilon_L,
\\
P^Q&=&P_Q+P_L.
\end{eqnarray}

In the hadronic phase, an improved isospin- and momentum-dependent interaction (ImMDI) model is used to describe the $\beta$-equilibrium and charge-neutral nuclear matter. In our previous study~\cite{Liu22}, the ImMDI model is fitted to the properties of cold symmetric nuclear matter (SNM), which is approximately reproduced by the self-consistent Greens function (SCGF) approach~\cite{Car14,Car18}  or chiral effective many-body pertubation theory ($\chi$EMBPT)~\cite{Wel15,Wel16}. And the parameters $x$, $y$ and $z$ are introduced to adjust the slope $L$ of symmetry energy, the momentum dependence of the symmetry potential, and the symmetry energy $E_{sym}(\rho_0)$ at saturation density, respectively. Recently, the discovery of GW170817 has triggered many analyses of neutron star observables to constrain nuclear symmetry energy. The average value of the slope parameter of the symmetry energy $L$ from the 24 new analyses of neutron star observables since GW170817 was about $L= 57.7 \pm 19$ MeV at a $68\%$ confidence level~\cite{BLi21}, which is consistent with the latest report of the slope parameter $L$ between 42 and 117 MeV from studying the pion spectrum ratio in heavy-ion collision in an experiment performed at RIKEN~\cite{Est21}. However, the Lead Radius Experiment (PREX-II) reported very recently new constraints on the neutron radius of $^{208}$Pb, which implies a neutron skin thickness of $R^{^{208}Pb}_{skin} = 0.283 \pm 0.071$ fm~\cite{Adh21} and constrains the slope parameter to $L= 106 \pm 37$ MeV~\cite{Ree21}, which is much larger than many previous constraints from microscopic calculations or experimental measurements~\cite{Tsa12,Lat13,BLi21}.  In order to better focus on the properties of quark matter, we thus choose for the hadronic phase a fixed parameter set, $x=-0.3$, $y=32$ MeV, and $z=0$, that would allow $2.08M_{\odot}$ neutron stars and still satisfy well nuclear matter constraints at saturation density, i.e., the binding energy $E_0(\rho_0) = -15.9$ MeV, the incompressibility $K_0 = 240$ MeV, the symmetry energy $E_{sym}(\rho_0) = 32.5$ MeV, the slope parameter $L=106$ MeV, the isoscalar effective mass $m^{\star}_s = 0.7m$, and the single-particle potential $U_{0,\infty} = 75$ MeV at infinitely large nucleon momentum.

The hadron-quark mixed phase is predicted to exist in the region between hadronic matter and quark matter based on various theoretical approaches. In the Maxwell construction, the coexisting hadronic and quark phases have equal pressure and baryon chemical potential but different electron chemical potential. The Gibbs construction is more generally adopted for the description of hadron-quark mixed phase, where the coexisting hadronic and quark phases are allowed to be charged separately. Besides, the mixed phase in the Gibbs construction persists within a limited pressure range, so it is convenient to form a massive neutron star containing the mixed phase. Both of the Maxwell and Gibbs constructions involve only bulk contributions, but the finite-size effects like surface and Coulomb contributions are neglected. The possible geometrical structure of the mixed phase has been extensively discussed in Refs.~\cite{Mar07,Na12,Wu19,Lug21,Ju21}. However, the large uncertainties in the structure and density range of the mixed phase are still present. In the present work, the hadron-quark mixed phase is described by imposing the Gibbs construction~\cite{GLe92,GLe01}:
\begin{eqnarray}
T^H&=&T^Q, \qquad \qquad P^H=P^Q,
\notag\\
\mu_B&=&\mu_B^H=\mu_B^Q,  \quad \mu_c=\mu_c^H=\mu_c^Q,
\end{eqnarray}
where $\mu_B$ and $\mu_c$ are the baryon and charge chemical potential, $P$ is the pressure, as well as the labels $H$ and $Q$ represent the hadronic and quark phases, respectively.
The Gibbs conditions for the chemical potentials can also be expressed as
\begin{eqnarray}
\mu_u&=&\frac{1}{3}\mu_n-\frac{2}{3}\mu_e,
\\
\mu_s&=&\mu_d=\frac{1}{3}\mu_n+\frac{1}{3}\mu_e.
\end{eqnarray}

Adding baryon number conservation, and charge neutrality conditions, the dense matter enters the mixed phase, in which the hadronic and the quark matter need to satisfy following equilibrium conditions:
\begin{eqnarray}
\mu_i&=&\mu_Bb_i-\mu_cq_i,  \quad P^H=P^Q,
\notag\\
\rho_B&=&(1-Y)(\rho_n+\rho_p)+\frac{Y}{3}(\rho_u+\rho_d+\rho_s),
\notag\\
0&=&(1-Y)\rho_p+\frac{Y}{3}(2\rho_u-\rho_d-\rho_s)-\rho_e-\rho_\mu,
\end{eqnarray}
where $Y$ is the baryon number fraction of the quark phase. The total energy density and pressure of the mixed phase are calculated according to
\begin{eqnarray}
\varepsilon^M&=&(1-Y)\varepsilon_H+Y\varepsilon_Q+\varepsilon_L,
\\
P^M&=&(1-Y)P_H+YP_Q+P_L.
\end{eqnarray}

The crust of hybrid stars, in our calculations, is considered to be divided into two parts: the inner and the outer crust as in the previous treatment~\cite{Jxu09C,Jxu09J}. The polytropic form $P = a + b\varepsilon^{4/3}$ has been found to be a good approximation to the inner crust EOS~\cite{Car03}, and the outer crust usually consists of heavy nuclei and electron gas, where we use the EOS in Ref.~\cite{Bay71}. Using the whole EOS from hadronic to quark phase, the mass-radius relation of hybrid stars can be obtained by solving the Tolman-Oppenheimer-Volkoff equation, which can be written as
\begin{eqnarray}
\frac{dP(r)}{dr}&=&-\frac{M(r)[\varepsilon(r)+P(r)]}{r^2}[1+\frac{4 \pi P(r)r^3}{M(r)}]
\notag\\
&\times&[1-\frac{2M(r)}{r}]^{-1},
\end{eqnarray}
where $\varepsilon(r)$ is the energy density and $P(r)$ is the pressure obtained from the equation of state. $M(r)$ is the gravitational mass inside the radius $r$ of the hybrid star given by
\begin{eqnarray}
\frac{dM(r)}{dr}&=&4 \pi r^2 \varepsilon(r).
\end{eqnarray}

The gravitational waves emitted from the merge of two compact stars are considered as another probe to the EOS of dense matter~\cite{Hin08,Rea09}. The tidal deformability $\Lambda$ of compact stars during their merger is related to the Love number $k_2$ through the relation $k_2 =3/2\Lambda\beta^5$~\cite{Hin08,Pos10}, which can be given by
\begin{eqnarray}
k_2 &=&\frac{8}{5}\beta^5(1-2\beta)^2[2-y_R+2\beta(y_R-1)]
\notag\\
&\times& \{2\beta[6-3y_R+3\beta(5y_R-8)]
\notag\\
&+& 4\beta^2[13-11y_R+\beta(3y_R-2)+2\beta^2(1+y_R)]
\notag\\
&+&3(1-2\beta)^2[2-y_R+2\beta(y_R-1)]\text{ln}(1-2\beta)\}^{-1},
\end{eqnarray}
where $\beta \equiv M/R $ is the compactness of the star, and $y_R \equiv y(R)$ is the solution at the compact star surface to the first
order differential equation
\begin{eqnarray}
r\frac{dy(r)}{dr}+y(r)^2+y(r)F(r)+r^2Q(r)=0,
\end{eqnarray}
with
\begin{eqnarray}
F(r) &=& \frac{r-4\pi r^3[\varepsilon(r)-P(r)]}{r-2M(r)},
\notag\\
Q(r)&=&\frac{4\pi r[5\varepsilon(r)+9P(r)+\frac{\varepsilon(r)+P(r)}{\partial P(r)/\partial \varepsilon(r)}-\frac{6}{4\pi r^2}]}{r-2M(r)}
\notag\\
&-&4[\frac{M(r)+4\pi r^3P(r)}{r^2(1-2M(r)/r)}]^2.
\end{eqnarray}
For a given central density $\rho_c$ and using the boundary conditions in terms of $y(0) = 2$, $P(0)=P_c$, $M(0)=0$ and $\epsilon(0)=0$, the mass $M$, radius $R$, and the tidal deformability $\Lambda$ of hybrid stars can be obtained once an complete EOS is supplied.

\section{Results and Discussions}
\label{RAD}

The aim of the present work is to analyze properties of quark-matter cores in massive hybrid stars. In the quark (3-flavor NJL) model, for the ease of discussions, we define the relative strength of the vector coupling, the scalar-isovector coupling, and the vector-isovector coupling respectively as $R_V=G_V/G_S$, $R_{IS}=G_{IS}/G_S$, and $R_{IV}=G_{IV}/G_S$. As is known, the position of the critical point for the chiral phase transition is sensitive to $R_V$~\cite{Asa89,Fuk08,Bra13}, which was later constrained within $0.5 <R_V <1.1$ from the relative elliptic flow $v_2$ splitting between protons and antiprotons as well as between $K^+$ and $K^-$ in relativistic heavy-ion collisions from the beam-energy scan (BES) program~\cite{Jxu14}. The strong vector-isovector interaction seems to be needed to reproduce the $v_2$ difference between $\pi^+$ and $\pi^-$ with the NJL transport approach~\cite{Liu19}. And the strength of $R_{IV}$ also leads to the isospin splittings of chiral phase transition boundaries, and thus affects the susceptibilities of conserved quantities~\cite{Liu21}. For the scalar-isovector interaction, it may result in a spinodal behavior in the EOS of the hadron-quark mixed phase and the corresponding hybrid star is unstable~\cite{Liu16}.  We thus mainly investigate, in this work, that the effect of vector and vector-isovector interactions on the quark matter in hybrid stars.

\begin{figure}[tbh]
\includegraphics[scale=0.33]{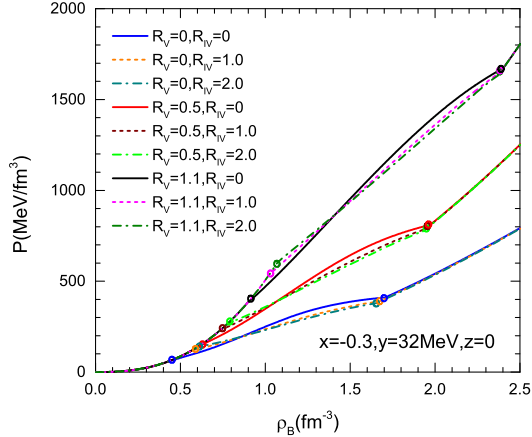}
\caption{(color online) The equation of states, pressure as a function of the baryon density, of hybrid star matter based on the ImMDI interaction for nuclear matter with a fixed parameter set ($x=-0.3$, $y=32$ MeV, and $z=0$) and the NJL model for quark matter with different coupling constants $R_V$ and $R_{IV}$. The two cycles with same color represent the the range of the hadron-quark mixed phase.} \label{fig1}
\end{figure}

\begin{figure*}[tbh]
\includegraphics[scale=0.42]{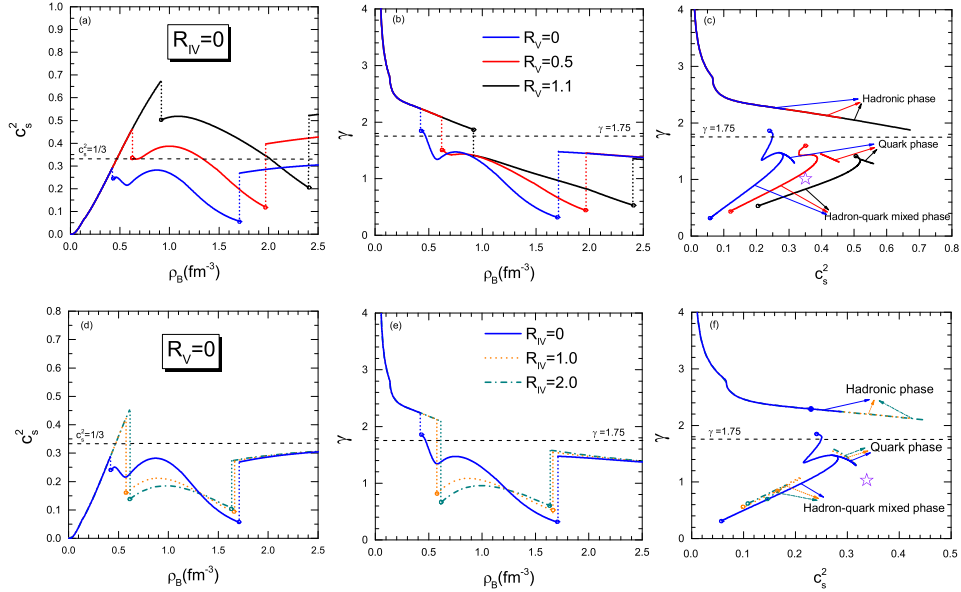}
\centering
\caption{(color online) The squared speed of sound $c_s^2$ and the polytropic index $\gamma$ as functions of the baryon density, as well as the relation between the polytropic index $\gamma$  and the squared speed of sound $c_s^2$  in hybrid star with the hadron-quark phase transition by varying the coupling constants $R_{V}$ (top panels) and $R_{IV}$  (bottom panels) from the NJL model. The dash lines $c_s^2=1/3$, and $\gamma=1.75$ as well as the violet star indicated the high-density conformal matter limit are also shown for comparison.}\label{fig2}
\centering
\label{EOS}
\end{figure*}

We first present in Fig.~\ref{fig1} the EOSs of hybrid star matter with the hadron-quark phase transition for different strength of the vector and vector-isovector interactions. The ImMDI interaction with a fixed parameter set, $x=-0.3$, $y=32$ MeV, and $z=0$, is used for nuclear matter, and the two cycles with same color in Fig.~\ref{fig1} represent the the range of the hadron-quark mixed phase. It can be seen the EOS is more sensitive to the strength of the vector interaction, and the EOS of hybrid star matter becomes stiffer with increasing vector strength $R_V$ for the quark matter. The vector-isovector interaction characterized by the coupling constant $R_{IV}$ slightly stiffens the EOS at low densities in the hadron-quark mixed phase, since its contribution is determined by the $G_{IV} (\rho_u -\rho_d)^2$ term in Eq. (2), which is sensitive to quark isospin asymmetry $\delta=(\rho_d-\rho_u)/\rho_B$. The effect of vector-isovector interaction gradually decreases, even leading to the equation of state being softer at high density in the mixed phase, which is due to the rapid decrease of isospin asymmetry with a larger $R_{IV}$~\cite{Liu22}. The isospin asymmetry of the $d$ and $u$ quark eventually decreases to zero at high densities, and thus the vector-isovector interactions have no effects on pure quark matter. In addition, as shown in Fig.~\ref{fig1}, with increasing coupling constants $R_V$ and $R_{IV}$ for the quark matter the onset of the phase transition is moving to higher densities since the transition pressure is also increasing under the Gibbs construction. It should be noted that the hadron-quark transition in most cases occurs at $2\rho_0\sim6\rho_0$ times saturation density, where the properties of quark matter are considered as an important factor affecting the maximum mass of hybrid stars.

Having the ensemble of EoSs with the hadron-quark phase transition for different strength of the quark interactions, we can determine the other properties of hybrid star matter. The quark matter at very high densities ($\rho_B\geq 40\rho_0$) is approximately scale-invariant or conformal, whereas in hadronic matter the degree of freedom is much smaller and the scale invariance is also violated by the breaking of chiral symmetry. These qualitative differences between hadronic and quark matter are reflected in the values taken by different physical quantities. The sound velocity $c_s$, which can be calculated from $c_s^2 = \partial{P}/\partial{\varepsilon}$,  takes the constant $c_s^2 =1/3$ in the exactly conformal matter corresponding to free massless fermions. However, the quantity in hadronic matter varies considerably: below saturation density, most hadronic models, such as chiral effective field theory (CET), indicate $c_s^2\ll1/3$, while at higher densities the square of maximum is predicted to be greater than 0.5~\cite{Gan09,Tew13}. On the other hand, the polytropic index $\gamma=d(\textrm{ln}P)/d(\textrm{ln}\varepsilon)$,  which is considered to be another good approximate criterion for the evidence of the quark-matter cores, has the value $\gamma = 1$ in conformal matter, while the hadronic models generically predict $\gamma\approx 2.5$ around and above saturation density~\cite{Kur10}.

\begin{figure}[tbh]
\includegraphics[scale=0.31]{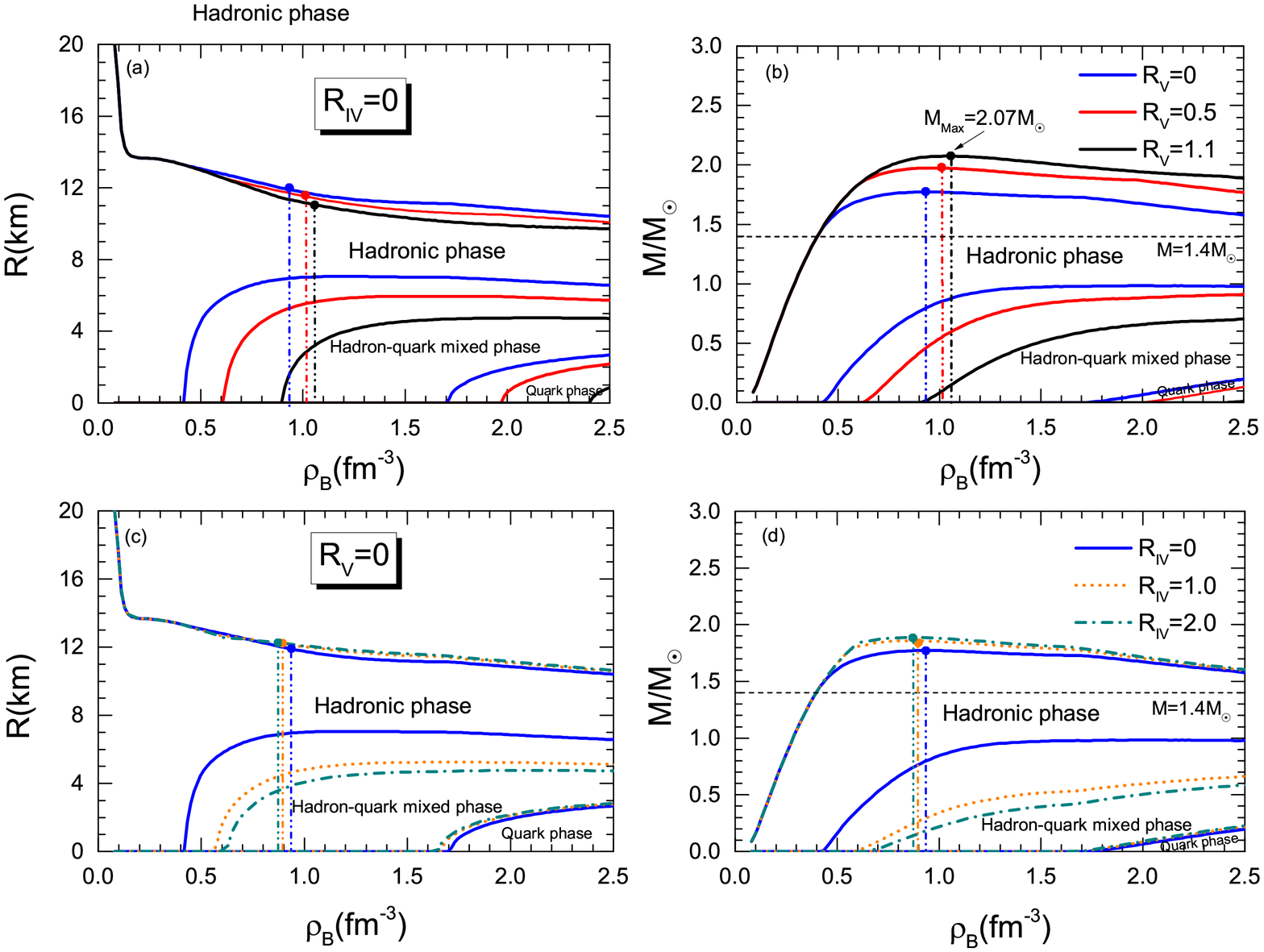}
\caption{(color online) The radii and mass of the hybrid stars containing quark phase, hadron-quark mixed phase, and hadronic phase as functions of baryon density based on the NJL model for quark matter with different coupling constants $R_V$ (upper panels) and $R_{IV}$ (lower panels). The different color solid points and dash-dotted vertical lines represent the maximum mass stars and corresponding baryon densities with different coupling constants, respectively.} \label{fig3}
\end{figure}

In Fig.~\ref{fig2}, we show that the squared speed of sound $c_s^2$ and the polytropic index $\gamma$ as functions of the baryon density in hybrid star with the hadron-quark phase transition by varying the coupling constants $R_{V}$ (top panels) and $R_{IV}$  (bottom panels) from the NJL model. It can be seen that at saturation density the squared speed of sound $c_s^2 =0.08$ and the polytropic index $\gamma=2.56$ from the ImMDI model with a fixed parameter set, $x=-0.3$, $y=32$ MeV, and $z=0$, is consistent with that in most hadronic models. And it should be noted that a step change of both the sound velocity and the polytropic index $\gamma$ occurs in the hadron-quark phase transition where the quarks appear and thus soften the EOS as a result of more degrees of freedom, and it is restored with the decrease of nucleon and lepton degrees of freedom in the high density quark phase. Also shown in panel (a) and (d) of Fig.~\ref{fig2} is the sound velocity $c_s^2 =1/3$ in the conformal limit, and it is seen that our results with a strong repulsive vector interaction for quark matter are larger than this limit at high densities, indicating that the corresponding EOS is stiffer than that of massless fermions. However, the results as shown in panel (b) and (e) of Fig.~\ref{fig2} with the vector-isovector interactions are quite different in quark phase, since the contribution of the vector-isovector interactions gradually decreases to zero at high densities. Meanwhile, we also found that $\gamma$ in the high density quark phase is insensitive to vector and vector-isovevtor interactions, and slowly approaches the value $\gamma = 1$ in conformal matter. In Fig.~\ref{fig2} (c) and (f), we show the relation between the polytropic index $\gamma$ and the squared speed of sound $c_s^2$ in hybrid star matter. The approximate rule following Ref.~\cite{Ann20} that the polytropic index $\gamma \leq 1.75$ can also be used as a criterion for separating hadronic from quark matter in our work. It should also be noted that the values of $\gamma$ and $c_s^2$  in the hadron-quark mixed phase are mostly less than 1 and 1/3 respectively, which may also be used as a new evidence for the existence of the hadron-quark mixed phase. Further, the step change of the sound velocity and the polytropic index in hadron-quark phase transition is relevant to the frequency of the main peak of the postmerger gravitational wave (GW) spectrum ($f_2$), which is expected to be confirmed by future kilohertz GW observations with third-generation GW detectors~\cite{Hua22}.

\begin{figure*}[tbh]
\includegraphics[scale=0.42]{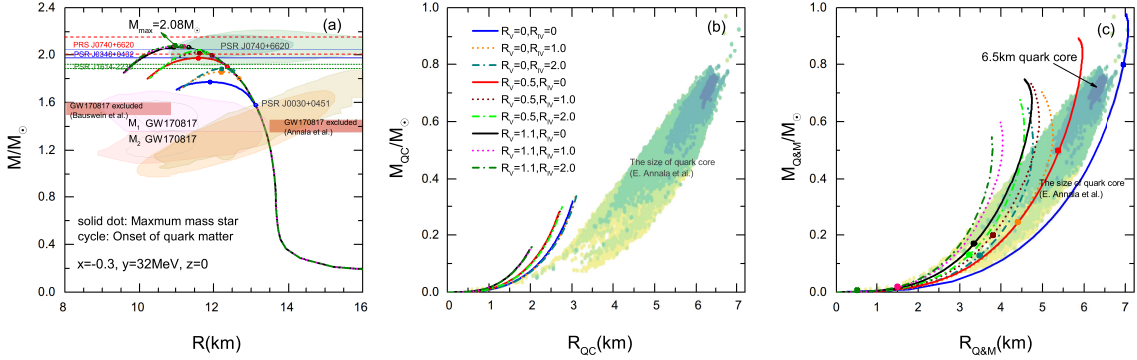}
\caption{(color online) Mass-radius relations of hybrid stars, pure quark cores (only including quark phase), and mixed cores (including quark phase and mixed phase) based on the NJL model for quark matter with different coupling constants $R_V$ and $R_{IV}$. The constraints from the bayesian analyses of the observational data from the pulsars PSR J0030+0451~\cite{Ril19,Mil19} and PSR J0740+6620~\cite{Ril21,Mil21}, and from the analyses of the gravitational wave signal from the neutron stars merger GW170817~\cite{Abb17} are shown in panel (a) by different color shaded regions. The horizontal bars represent the mass measurements of PSR J1614-2230 ($1.908\pm0.016M_{\odot}$)~\cite{Dem10,Fon16,Arz18}, PSR J0348+0432 ($2.01\pm0.04M_{\odot}$)~\cite{Ant13}, and PSR J0740+6620 ($2.08\pm0.07M_{\odot}$)~\cite{Cro20,Ril21,Mil21}. The two red bands correspond to excluded regions derived from GW170817 observations~\cite{Bau17,Ann18} are also shown in panel (a) for comparison. Furthermore, the shaded regions in panels (b) and (c) show the size of quark cores from Ref.~\cite{Ann20} for comparison. The different color solid dots in panel (a) and panel (c) represent the maximum mass stars with different coupling constants, while the cycles represent the onset of quark matter.} \label{fig4}
\end{figure*}

 Essentially all available equation of states can be used to predict the radius and mass information of compact stars. As shown in Fig.~\ref{fig3}, we show the radii and mass of the hybrid stars containing quark phase, hadron-quark mixed phase, and hadronic phase as functions of baryon density based on the NJL model for quark matter with different coupling constants $R_V$ and $R_{IV}$. The results shown in the upper panels indicate that both the mass and radii of quark phase, mixed phase and the complete hybrid star are sensitive to the strength of the vector interaction. Qualitatively, the radii of both quark phase, mixed phase and the complete hybrid star decrease with the increment of $R_V$, which is due to the fact that quarks appear later by the increase of $R_V$. Although the similar dependence occurs in the mass of the quark phase and mixed phase, the mass of complete star increases with the increase of $R_V$, since the mass of complete star mainly depends on the stiffness of the equation of state. However, a larger $R_V$ also means that the fraction of quark matter is smaller in the core of hybrid star. Particularly, if $R_V$ is large enough, the onset density of quark matter will be larger than the central density of the star, and thus no pure quark-matter core can appear in the hybrid stars. By contrast, as shown in the lower panels, the vector-isovector interactions make more contribution to the hadron-quark mixed phase, but have slight effects on the mass and radii of quark phase and the complete hybrid star. In addition, it should be noted that the maximum mass of hybrid stars constrains mostly the EOS of hybrid star matter at $2 \sim5$ times saturation density, which is related to the onset of the hadron-quark transition, and thus the properties of mixed phase will affect the maximum mass of hybrid stars. Although decreasing $R_V$ and $R_{IV}$ can soften the EOS of quark matter and thus decrease the maximum mass of hybrid stars, the softening of the quark matter EOS also makes the onset of first order phase transition appear earlier, so that the mass and radius of the mixed phase contribute more to that of the complete star. After the mass of hybrid stars reaches the maximum value, further increasing the central density will reduce both the maximum mass and the radius, making the hybrid star unstable. It is clearly seen in Fig. 3 (b) and (d) that there is no pure quark matter in the inner core of the stable hybrid stars. Furthermore, we also can see that properties of quark matter have no effect on the $M = 1.4M_{\odot}$ hybrid star as a result of no quark matter (including pure quark matter and mixed phase matter) inner core, which can also be confirmed by the criterion of the polytropic index, since we find that the polytropic index for neutron stars with $M = 1.4M_{\odot}$ (corresponding to $\rho_B=0.4$fm$^{-3}$) always satisfies $\gamma = 2.25$, implying that the stars are composed of hadronic matter as expected.

\begin{figure*}[tbh]
\includegraphics[scale=0.42]{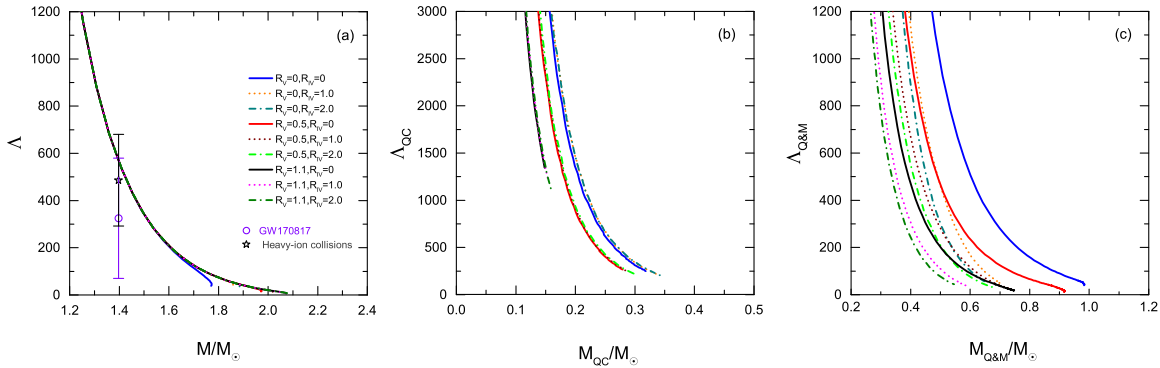}
\caption{(color online) Relations between the dimensionless tidal deformability and the mass of hybrid stars, pure quark cores, and mixed cores based on the NJL model for quark matter with different coupling constants $R_V$ and $R_{IV}$. The error bars at $1.4M_{\odot}$ correspond to the constraints on the tidal deformability $70 \leq \Lambda \leq 580$ based on the improved analyses of GW170817 by LIGO and Virgo Collaborations~\cite{Abb18} as well as the prediction of $292 \leq \Lambda \leq 680$ from heavy-ion collisions~\cite{BLi06}.}\label{fig5}
\end{figure*}

We also show in Fig.~\ref{fig4}  the mass-radius relations of hybrid stars based on the NJL model for quark matter with different coupling constants $R_V$ and $R_{IV}$. Constraints from multi-messenger astronomy observations are shown in the panel (a) by the different color regions/bands~\cite{Abb17,Ril19,Mil19,Ril21,Mil21,Bau17,Ann18} and horizontal bars~\cite{Dem10,Fon16,Arz18,Ant13,Cro20,Ril21,Mil21}. It can be seen that the observed maximum mass of hybrid stars depend on the quark coupling constants $R_V$ and $R_{IV}$. The results of hybrid stars in all parameter sets are mostly consistent with the constraints from the pulsars PSR J0030+0451 and the neutron stars merger GW170817. Except for the case $R_V=0$ and $R_{IV}=0$, the maximum values of other curves are close to the measurements of massive neutron stars ($\approx 2M_{\odot}$),  Especially in the case $R_V=1.1$ and $R_{IV}=2.0$, the maximum mass of hybrid stars $M_{\textrm{max}}=2.08M_{\odot}$ is consistent with the measurement of PSR J0740+6620, which is considered as the highest reliably determined compact star mass. The radii of hybrid stars are known to be determined by the pressure at densities around $\rho_0\sim2\rho_0$ ~\cite{Lat00,Lat01},  and thus constrain mostly the nuclear matter but is not sensitive to quark matter. The hybrid star matter EOS consists of charge-neutral matter in $\beta$-equilibrium that has a hadron-quark phase transition from hadronic to quark matter.

 To better understand the effects of quark matter on the hybrid stars, the relations between the quark core mass $M_{QC}$ and radius $R_{QC}$ as well as between the mixed core mass $M_{Q\&M}$ and radius $R_{Q\&M}$ are shown in Fig. 4 (b) and (c). The $M-R$ relations of the quark matter cores can be determined using the TOV equations by integrating them from the central baryon density to three distinct points: (1) the pressure of onset of pure quark phase where $r=R_{QC}$ and $M(r)=M_{QC}$, (2) the pressure of onset of mixed phase where $r=R_{Q\&M}$ and $M(r)=M_{Q\&M}$, and (3) zero pressure where $r=R$ and $M(r)=M$. We also obtained the tidal deformabilities of the quark matter cores using the same approach. It is clearly seen from the two right panels of Fig. 4 that the coupling constants $R_V$ and $R_{IV}$ have a competitive effect: while the coupling constants increase the hybrid star mass up to $2.08M_{\odot}$, they also decrease the mass and radius of the quark core and the mixed core. The size of quark cores from Ref.~\cite{Ann20} is also shown by the shaded regions in Fig. 4 (b) and (c) for comparison. One should be noted that the shaded regions are different from that in panel (a), which are not derived from observational results, but statistical analysis based on the results of model-independent calculations proposed in ref.~\cite{Ann20}. It can be seen from the figure that the mass and radius containing the mixed phase are more approaching to the results of Ref.~\cite{Ann20}. As previously mentioned, once the mass of the hybrid star reaches its maximum value, the hybrid star becomes unstable, and no pure quark matter is found in the inner core of stable hybrid stars. The maximum mass and radius of the quark matter core in a stable hybrid star can even reach $0.80M_{\odot}$ and 6.95 km, respectively, which are close to half of the maximum mass and radius of the complete star. For the hybrid star matter satisfying the requirement of supporting a $2M_{\odot}$  stable star, the maximum radius of the quark matter core is $R_{Q\& M}\approx3.85$ km.

After the GW170817 event, much efforts have been devoted to constraining the EOS or related model parameters by comparing various calculations with the range of tidal deformability $70 \leq \Lambda_{1.4} \leq 580$ from the improved analyses reported by LIGO and Virgo Collaborations. The measurements of the tidal deformability of compact stars constrain not only the EOS of dense nuclear matter but also the fundamental strong interactions of quark matter. The relations between the tidal deformability and the mass using NJL model by varying the coupling constants $R_V$ and $R_{IV}$ are shown in Fig.~\ref{fig5}. For comparison, we display the error bars at $1.4M_{\odot}$ in the panel (a), which correspond to the constraints on the tidal deformability $70 \leq \Lambda \leq 580$  based on the improved analyses of GW170817 by LIGO and Virgo Collaborations~\cite{Abb18} as well as the prediction of $292 \leq \Lambda \leq 680$ from heavy-ion collisions~\cite{BLi06}. We can see that $\Lambda$ decreases rapidly as the mass of the neutron star increases. This is due to the factor that given the smaller range of allowed radii for larger massive stars, the spread in the tidal deformability is also naturally much tighter than for lower-mass neutron stars. The results in panel (a) show that the vector and vector-isovector interactions have slightly effects on the minimum deformability which is related to the difference in maximum mass of hybrid stars. For a given mass $M =1.4 M_{\odot}$, the deformability $\Lambda_{1.4}$ increases with increasing radius of hybrid star, and thus these quark interactions have no effect on the $\Lambda_{1.4}$ of hybrid stars, as expected. However, for the massive hybrid star, the couplings have effect on the quark matter EOSs and lead to the difference of the tidal deformability of quark-matter cores. This is illustrated in panels (b) and (c) of Fig.~\ref{fig5} which display the dimensionless tidal deformability of the pure quark cores and the mixed cores. Generally, a stiffer EOS will lead to a compact star with the larger mass and radius, and thus increasing the tidal deformability. In contrast, in the quark core and mixed core, the mass and radius of the core decreases due to the stiffening of the equation of state by the coupling constants, which also decreases the tidal deformability instead.

\section{Summary and Outlook}
\label{viscosity}
 In this work, we have investigated the properties of quark matter core by using a hybrid star with the hadron-quark phase transition. The quark matter interactions in hybrid stars are described based on 3-flavor NJL model by varying vector and vector-isovector coupling constants. In conclusion, we found that the hybrid star matter EOS is more sensitive to the strength of the vector interaction, and the EOS of hybrid star matter becomes stiffer with increasing vector strength $R_V$. The vector-isovector interaction characterized by the coupling constant $R_{IV}$ make main contribution to the hadron-quark mixed phase. In the study of other properties of hybrid star matter, it should be noted that a step change of both the sound velocity and the polytropic index $\gamma$ occurs in the hadron-quark phase transition, and it is restored with the decrease of nucleon and lepton degrees of freedom in the high density quark phase. The approximate rule following Ref.~\cite{Ann20} that the polytropic index $\gamma\leq 1.75$ can also be used as a criterion for separating hadronic from quark matter in our work.  Using the hybrid star matter EOS, we predict the radius and mass information of quark-matter cores inside hybrid stars. Although the coupling constants increases the hybrid star maximum mass up to $2.08M_{\odot}$, they also decrease the mass and radius of the quark core and the mixed core. With different quark coupling constants, we also found that the maximum mass and radius of the quark matter core in a stable hybrid star can reach $0.80M_{\odot}$ and 6.95 km, which are close to half of the maximum mass and radius of the complete star. However, properties of quark matter have no effect on the $M = 1.4M_{\odot}$ compact star as a result of no quark matter inner core, which can also be confirmed by the criterion of the polytropic index, and thus the quark interactions have no effect on the tidal deformability $\Lambda_{1.4}$ of hybrid stars. The coupling constants $R_V$ and $R_{IV}$ in the NJL model determine the EOS of dense matter and also affect the critical point as well as the QCD phase structure. To further explore the QCD phase structure and search for the signal of the critical point between the crossover and the first-order transition, experimental programs such as the beam-energy scan (BES) at RHIC and the compressed baryonic matter (CBM) at Facilities for Antiproton and Ion Research (FAIR) were proposed. The promising results are available to provide more constraints on the EOSs of quark matter, which are helpful in the understanding of the QCD phase structure and properties of hybrid stars. In addition, some of the new discoveries and observations in astrophysics also provide more rigorous constraints on the QCD phase structure. The binary-neutron-star (BNS) merger simulations, for example, indicate that the sudden decrease in the gravitational wave frequency is closely related to the hadron-quark phase transition, which is expected to be confirmed by future kilohertz GW observations with third-generation GW detectors.
\notag\\
\begin{acknowledgments}
This work is supported by the National Natural Science Foundation of China under Grants No. 12205158 and No. 11975132, and the Shandong Provincial Natural Science Foundation, China Grants No. ZR2021QA037, No. ZR2022JQ04, No. ZR2019YQ01, and No. ZR2021MA037.
\end{acknowledgments}

\end{document}